\pdfoutput=1

\documentclass{appolb}

\usepackage{amssymb}
\usepackage{hyperref}
\usepackage{graphicx}
\usepackage{amsmath}
\usepackage[numbers,sectionbib,sort&compress]{natbib}

\newcommand{\vect}[1]{\boldsymbol{#1}}

\begin{document}
\title{Double parton distributions of the pion%
\thanks{Talk presented by WB at {\em Excited QCD 2020}, Krynica Zdrój, Poland, 2-8 February 2020}
\thanks{Supported by the Polish National Science Centre (NCN)
Grant 2018/31/B/ST2/01022, the Spanish Ministerio de Economia y
Competitividad and European FEDER funds Grant FIS2017-85053-C2-1-P,
and Junta de Andaluc\'{\i}a Grant FQM-225.}}

\author{Wojciech Broniowski$^{1,2}$\thanks{Wojciech.Broniowski@ifj.edu.pl}, Enrique Ruiz Arriola$^3$\thanks{earriola@ugr.es}
\address{$^{1}$Institute of Physics, Jan Kochanowski University, PL-25406~Kielce, Poland}
\address{$^{2}$The H. Niewodnicza\'nski Institute of Nuclear Physics, \\ Polish Academy of Sciences, PL-31342~Cracow, Poland}
\address{$^{3}$Departamento de F\'isica At\'omica, Molecular y Nuclear \\ and Instituto Carlos I de F\'{\i}sica Te\'orica y Computacional,
                   Universidad de Granada, E-18071 Granada, Spain}
}

\maketitle

\abstract{
We present a calculation of valence double parton distributions of the
pion in the framework of chiral quark models.  The result obtained at
the low-energy quark model scale is particularly simple, where in the
chiral limit a factorized form follows, $D(x_1,x_2, \vec{q})
= \delta(1-x_1-x_2) F(\vec{q})$ with $x_{1,2}$ standing for the
longitudinal momentum fractions of the valence quark and antiquark,
and $\vec{q}$ denotes the relative transverse momentum.  For
$\vec{q}=\vec{0}$ the result satisfies the Gaunt-Sterling sum rules.
The QCD evolution to higher scales is carried out within the dDGLAP
framework. We argue that the ratios of the valence Mellin moments
$\langle x_1^n x_2^m \rangle / \langle x_1^n \rangle \langle
x_2^m \rangle $, which do not depend on the dDGLAP evolution, provide
particularly convenient measures of the longitudinal correlations
between the partons. Such ratios could be probed in future lattice QCD
simulations.  }

\bigskip
\bigskip

This contribution is based on our recent work on double parton distributions of the pion~\cite{BW-ERA-LC2019,Broniowski:2019rmu}, also 
explored by Courtoy et al.~\cite{Courtoy:2019cxq}.
The old story of double parton distribution
(dPDFs)~\cite{Kuti:1971ph}, followed with early experimental searches
by the Axial Field Spectrometer Collaboration at the CERN
ISR~\cite{Akesson:1986iv} and by the CDF Collaboration at
Fermilab~\cite{Abe:1993rv,Abe:1997xk}, has recently picked up renewed
interest~\cite{Bartalini:2011jp,Snigirev:2011zz,Luszczak:2011zp,Manohar:2012jr,Manohar:2012pe,d'Enterria:2012qx,Bartalini:2017jkk}
with a growing evidence from the LHC,
e.g.,~\cite{Aad:2013bjm,Sirunyan:2019zox}. Experimentally, the results
concern mostly the structure of the proton
which naturally provides a stable target . The
proton, however, is much harder to model theoretically than the pion,
for which there is lack of experimental data. On the other hand, the pion is the simplest hadronic bound state appearing as a
pseudo-Goldstone boson of the spontaneously broken chiral symmetry.
Hence we explore the pion, with the hope the result can be verified in
the near future with lattice QCD studies, where the pion is readily
available at the physical quark masses.

\begin{figure}
\begin{center}
\includegraphics[angle=0,width=0.35\textwidth]{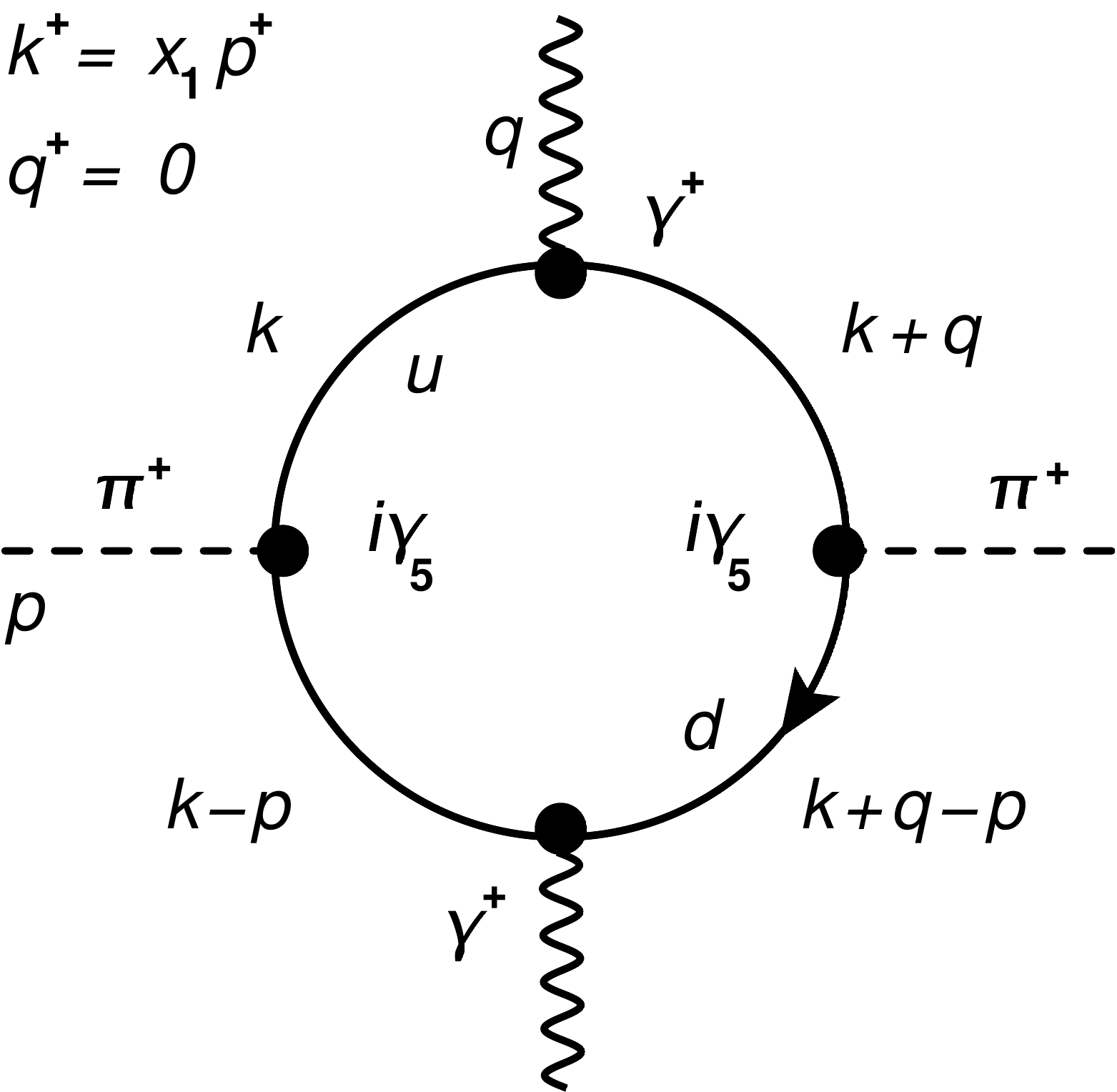} 
\end{center}
\caption{Diagram for evaluation of the double valence quark distributions of $\pi^+$ at the leading-$N_c$ order (one-loop) in the NJL model.
Note that $q^+=0$, whereas integration over $q^-$ is carried out.
\label{fig:diag}} 
\end{figure}

The field-theoretic definition of valence dPDFs generalizes the case of the single parton distribution functions (sPDFs)  
namely  (see \cite{Diehl:2010dr} and references therein)
\begin{eqnarray}
&&  \hspace{-2mm} D_{j_{1} j_{2}}(x_1,x_2,\vect{b})
 =  2 p^+ \int d b^-\,
        \frac{d z^-_1}{2\pi}\, \frac{d z^-_2}{2\pi}\;
          e^{i ( x_1^{} z_1^- + x_2^{} z_2^-) p_{}^+} \nonumber \\
 && ~~~~ \times
    \langle p |\, {\mathcal{O}_{j_1}(b,z_1)\, \mathcal{O}_{j_2}(0,z_2)}
    \,| p \rangle
    \bigl|_{z_1^+ = z_2^+ = b_{\phantom{1}}^+ = 0\,,
    \vect{z}_1^{} = \vect{z}_2^{} = \vect{0}}, \label{eq:defd}
\end{eqnarray}
where $p$ is the momentum of the hadron, $j_{1,2}$ labels the quark species, $x_{1,2}$ are the fractions of the light-cone momentum of the 
hadron carried by the valence quark or antiquark,  $z_{1,2}$ are the coordinates on the light front, $b$ is the spatial separation of the two bilocal operators
for the valence quark and antiquark,
\begin{eqnarray}
&& \hspace{-7mm} \mathcal{O}_{q}(y, z) = \tfrac{1}{2}\, \bar{q} ( y- \tfrac{z}{2}) \gamma^+  q ( y+ \tfrac{z}{2} ), 
\mathcal{O}_{\bar{q}}(y, z) =-\tfrac{1}{2}\, \bar{q} ( y + \tfrac{z}{2}) \gamma^+  q ( y - \tfrac{z}{2}).
\end{eqnarray} 
Our convention for the light-cone variables is $a^\pm = (a^0 \pm a^3) /\sqrt{2}$, and $\vect{a}=(a_1,a_2)$.
In the applied chiral quark model the evaluation of (\ref{eq:defd}) is carried out in the Fourier-conjugated space, as illustrated by the 
diagram in Fig.~\ref{fig:diag}, where for definiteness we take the case of the charged pion, $\pi^+$.
We note that the momentum $q$  flows between the two probing operators, bringing in the information on transverse structure of the pion. The integration over 
$q^-$ enforces the constraint $b^+=0$, whereas the transverse component $\vec{q}$ is the Fourier-conjugate
variable corresponding to $\vec{b}$.

In the chiral limit the result is particularly simple~\cite{BW-ERA-LC2019,Courtoy:2019cxq,Broniowski:2019rmu}:
\begin{eqnarray}
D_{u\bar{d}}(x_1,x_2,\vect{q})= 1 \times \delta(1-x_1-x_2) \Theta   F(\vec{q}), \label{eq:dp}
\end{eqnarray}
where the presence of the $\delta$ function reflects the conservation of the light-front momentum and $\Theta$ reflects the proper support 
$0 \le x_1,x_2 \le 1$, following from the Lorentz symmetry preserved in the  calculation.  The form factor in the transverse momentum,
$ F(\vec{q})$, depends on the adopted low-energy regularization scheme (see~\cite{Broniowski:2019rmu}), which 
is a necessary ingredient of the effective model. We note that the factorization of the light-front and transverse dynamics holds 
exactly only in the strict chiral limit.  We note that, importantly, Eq.~(\ref{eq:dp}) satisfies the Gaunt-Stirling (GS) sum rules~\cite{Gaunt:2009re},
holding in the case of $\vec{q}=0$ and relating integrals of dPDFs with sPDFs. 
As the GS sum rules are preserved by the QCD evolution, they hold at any scale in the applied framework.

\begin{figure}
\begin{center}
\resizebox{0.49\columnwidth}{!}{\includegraphics{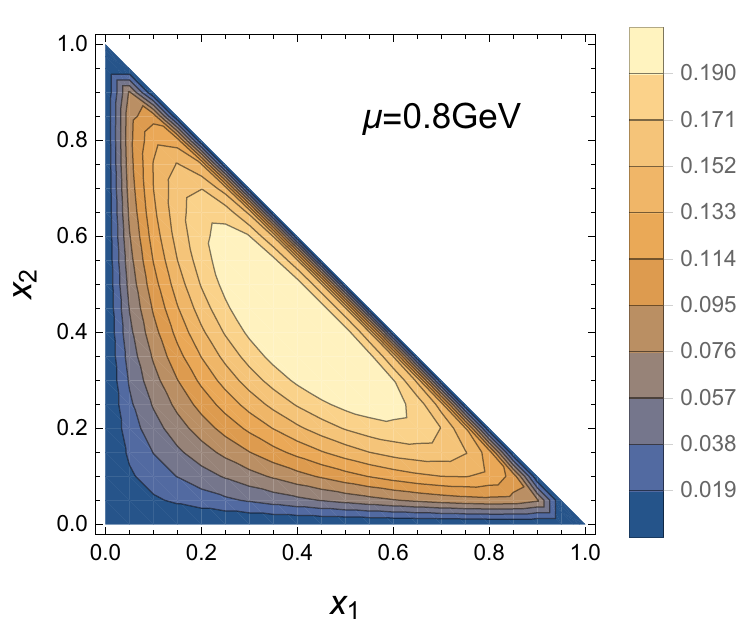}}       \resizebox{0.49\columnwidth}{!}{\includegraphics{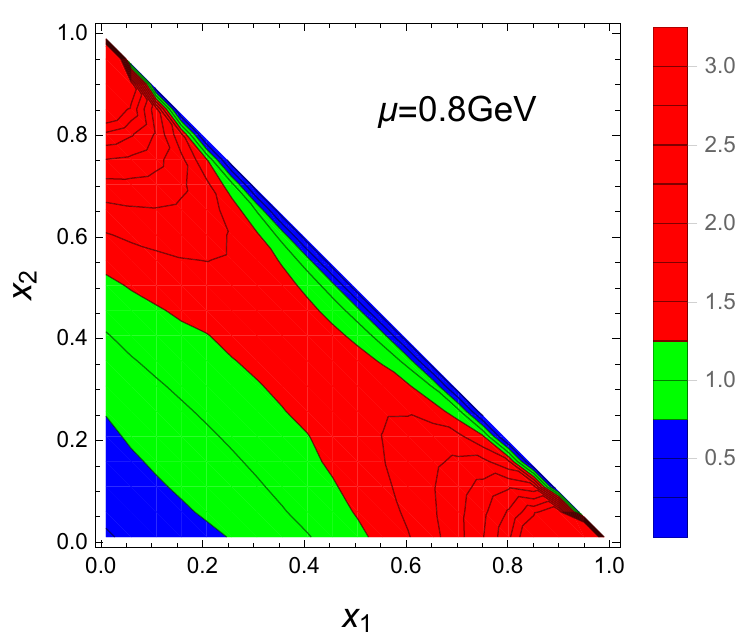}} \\
\resizebox{0.49\columnwidth}{!}{\includegraphics{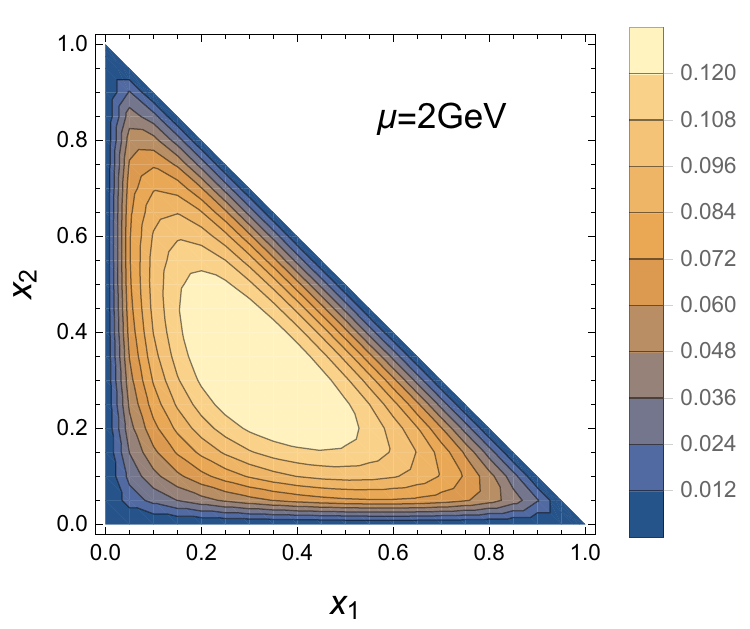}}         \resizebox{0.49\columnwidth}{!}{\includegraphics{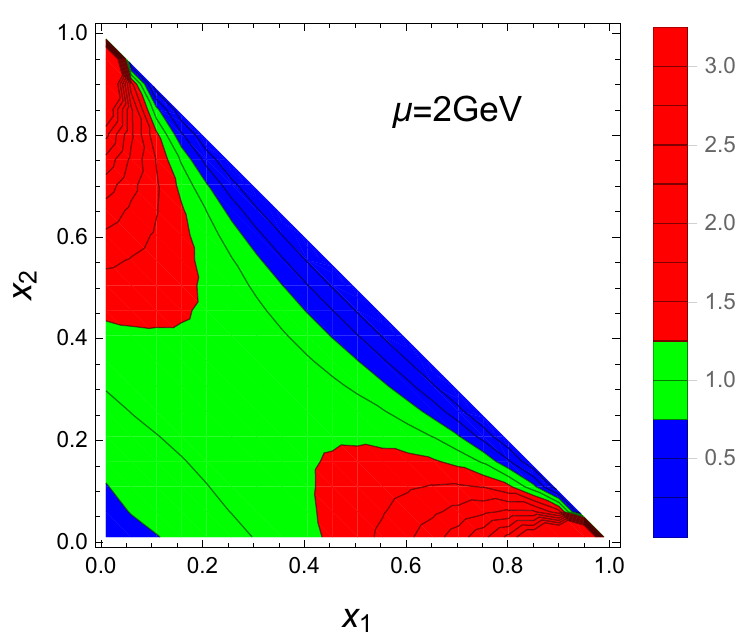}} \\
\resizebox{0.49\columnwidth}{!}{\includegraphics{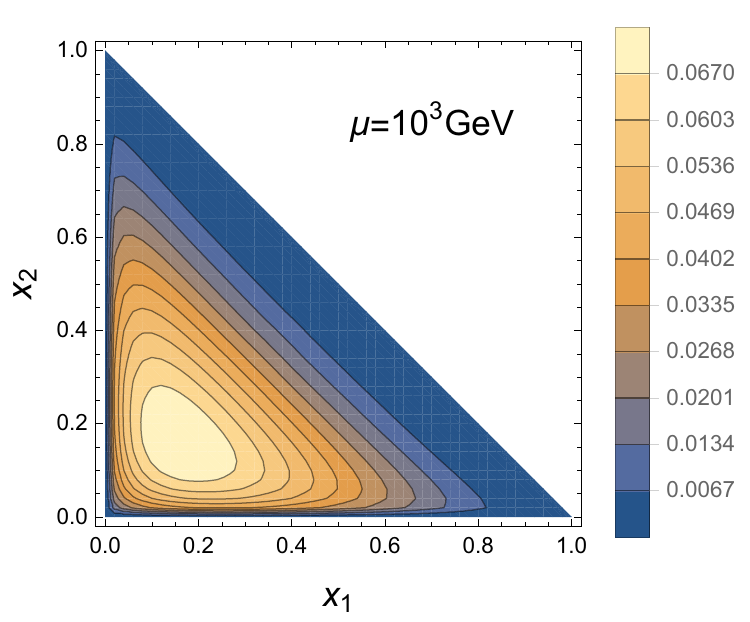}} \resizebox{0.49\columnwidth}{!}{\includegraphics{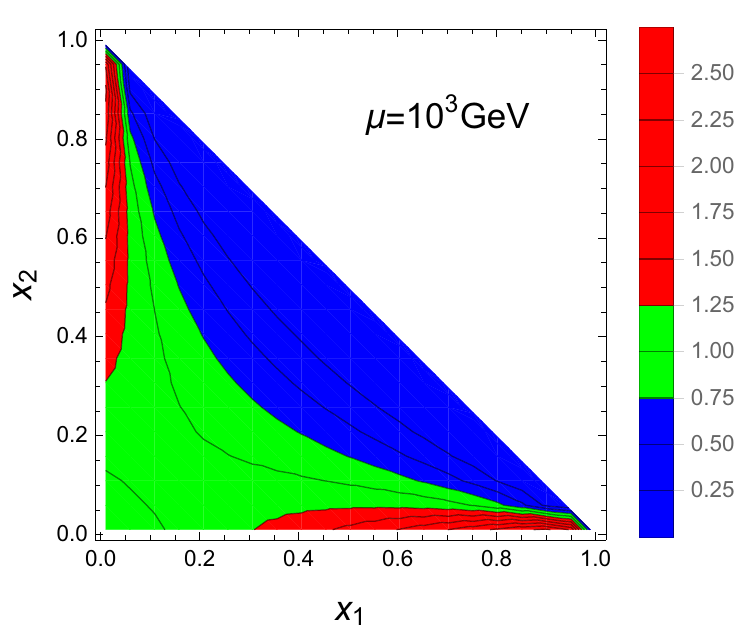}} 
\end{center}
\caption{Left: Valence dPDF of the pion, $x_1 x_2 D_{\rm u \bar{d}}(x_1,x_2)$, evolved with the dDGLAP equations 
to subsequent scales $\mu$. 
Right: Corresponding correlation $D_{\rm u \bar{d}}(x_1,x_2)/D_u(x_1)D_{\bar{d}}(x_2)$. \label{fig:dev}}
\end{figure}

The QCD evolution is a necessary ingredient of the scheme, as first noticed in the study of sPDFs of the pion by Davidson and one of us (ERA)~\cite{Davidson:1994uv}.
Since then the model, with evolution discussed in detail in~\cite{Broniowski:2007si}, has been applied to numerous soft matrix elements relevant for high-energy 
processes, such as 
the parton distribution amplitude (PDA)~\cite{RuizArriola:2002bp}, the generalized parton distributions (GPD)~\cite{Broniowski:2007si}, 
and the quasi parton distributions (QDF)~\cite{Broniowski:2017wbr,Broniowski:2017gfp}, with successful outcome.
The dDGLAP evolution scheme~\cite{Kirschner:1979im,Shelest:1982dg} is straightforward to implement in terms of the Mellin moments, similarly
to sPDFs~\cite{Broniowski:2013xba,Broniowski:2019rmu}.
Our results after evolution are shown in the left panels of Fig.~\ref{fig:dev}.
The right panels display the correlation $D_{\rm u \bar{d}}(x_1,x_2)/D_u(x_1)D_{\bar{d}}(x_2)$. The regions in the plots marked with green have the 
correlation within 25\% from the unity. We note that increasing the evolution scale brings the correlation  closer to unity (cf. a similar behavior 
fond  for the gluon distributions of the nucleon in~\cite{Golec-Biernat:2015aza}). 

A very simple measure of correlation is the ratio
${\langle  x_1^n x_2^m \rangle}/{\langle  x_1^n \rangle \langle x_2^m \rangle}$
which is  independent of the evolution scale. In our model (in the chiral limit) we have
\begin{eqnarray}
\frac{\langle  x_1^n x_2^m \rangle}{\langle  x_1^n \rangle \langle x_2^m \rangle}= \frac{(1+n)!(1+m)!}{(1+n+m)!}. \label{eq:momnjl}
\end{eqnarray}
The lowest correlation ratios can hopefully be probed in the future lattice QCD simulations.

In summary, our results from a covariant framework combining chiral quark model evaluation with QCD evolution  
satisfy all formal requirements, in particular the GS sum rules. In the chiral limit,  the light-front and transverse dynamics is
factorized, whereas in the $x_1$ and $x_2$ spaces there is no factorization due to the momentum conservation. 
We have proposed simple scale independent ratios of the  Mellin moments as probes of the correlation. Lowest ratios could be 
obtained from future lattice simulations.

\bibliography{dPDF,refs,covid-19}

\end{document}